\documentclass{jps-cp}
\usepackage{amsmath}
\usepackage{amssymb}
\usepackage[mathscr]{euscript}
\usepackage{graphicx}					
\usepackage[crop=pdfcrop,process=auto,cleanup={.dvi,.ps,.pdf,.log}]{pstool}
\renewcommand{\vec}[1]{{\pmb #1}}
\newcommand{\grad}{\pmb\nabla}
\def\Tr#1{\mbox{Tr}\big\{#1\big\}}
\title{Disorder Induced Anomalous Thermal Hall Effect in Chiral Phases of Superfluid $^3$He}
\author{Priya \textsc{Sharma}$^{1}$, Anton \textsc{B. Vorontsov}$^{2}$ and J.A. \textsc{Sauls}$^{3}$}
\inst{$^{1}$Department of Physics, Royal Holloway University of London, Egham, Surrey, UK \\
$^{2}$Department of Physics, Montana State University, Bozeman, MT, USA \\
$^{3}$Department of Physics, Northwestern University, Evanston, IL, USA}
\email{priya.sharma@rhul.ac.uk}

\recdate{Aug 7, 2022}

\abst{
NMR experiments on liquid $^3$He infused into uniaxially anisotropic silica aerogels show the stabilisation of two equal-spin-pairing chiral phases on cooling from the normal phase. The alignment of the chiral axis relative to the anisotropy axis for these phases is predicted to depend upon temperature. A chiral A-like phase is also stabilized when $^3$He is confined to a slab of thickness $D\sim \xi$, the superfluid coherence length. For both types of confinement, scattering of quasiparticles by the random potential - aerogel or surface - is pair breaking and generates a sub-gap density of quasiparticle states. The random field also conspires with the chiral order parameter to generate skew scattering of quasiparticles in the plane normal to the chiral axis. This scattering mechanism leads to anomalous thermal Hall transport for nonequilibrium quasiparticles driven by a thermal gradient. We report theoretical results for the anomalous thermal Hall conductivity for theoretical models for chiral phases of $^3$He in both anisotropic aerogel and slabs. The anomalous thermal Hall effect (ATHE) provides an important tool to identify signatures of broken time-reversal and mirror symmetries and topology in chiral superconductors/superfluids. 
}
\kword{Superfluid $^3$He, slab, films, confinement, chirality, thermal Hall effect}
\begin{document}
\maketitle

\vspace*{-15mm}
\section{Introduction}

NMR experiments on liquid $^3$He infused into uniformly anisotropic uniaxially compressed high-porosity silica aerogels~\cite{pol08} have been interpreted to show the stabilisation of an equal spin pairing(ESP) ABM~\cite{and61} phase on cooling from the normal phase~\cite{pol12}. The chiral axis, $\hat{\vec\ell}$ of this A- phase is identified to be aligned parallel to the strain axis, $\hat{\vec z}$ of the aerogel. Scattering of quasiparticles off the aerogel matrix modifies the density of states (DOS) in the $^3$He-A - aerogel system generating gapless states associated with pairbreaking effects of this scattering~\cite{sha03}. Quasiparticles with this modified DOS can be probed through their response to external perturbations. On the application of a thermal gradient, anomalous thermal Hall transport arises in the orbital plane, due to the coupling of the $p$-wave perturbation, $\vec{v}_{\vec{p}}\cdot\grad T$ where $\vec{v}_{\vec{p}}$ is the velocity of quasiparticles of momentum $\vec{p}$, to the chiral $p$-wave order parameter, both belonging to the $l=1$ channel
in orbital space.

The order parameter for the $^3$He-A phase is given by ${\hat{\Delta}}_{\alpha j} = \Delta_0\,\vec{d}_{\alpha} (\hat{\vec p}.\hat{\vec m} + i\,\hat{\vec p}.\hat{\vec n})_j$ where $\hat{\vec\ell} = \hat{\vec m}\times\hat{\vec n}$ is the chiral axis.
For $^3$He-A confined to a slab of thickness $D$, with the chiral axis $\hat{\vec\ell}$ aligned perpendicular to the slab surfaces, non-specular scattering of quasiparticles off the confining surfaces generates branch conversion scattering and pair breaking, the latter leading to a modified DOS with sub-gap states penetrating the thin film. Branch conversion scattering of non-equilibrium quasiparticles by the chiral order parameter leads to anomalous thermal Hall transport for heat currents in the plane of the slab and perpendicular to the chiral axis.

\section{Theoretical Formalism}

We employ the Keldysh formulation~\cite{kel65} of quasiclassical theory~\cite{eil68,lar69,ser83} which describes phenomena that occur on characteristic length scales much larger than the Fermi wavelength, $k_f^{-1}$ and characteristic time scales much longer than the inverse Fermi energy $\varepsilon_f^{-1}$ (Here, $\hbar = 1$). The heat current density $\vec{j}_{\varepsilon}$ is given by
\begin{equation}
\vec{j}_{\varepsilon}=N_f\int\nolimits\frac{d\Omega_{\hat{\vec p}}}{4\pi}
                         \int\nolimits\frac{d\varepsilon}{4\pi i}\,(\varepsilon\,\vec{v}_{\vec{p}})\,
			 \Tr{\widehat{g}^K(\vec{p},\varepsilon)}
\,,
\label{je}
\end{equation}
where $N_f$ is the normal-state quasiparticle density of states at the Fermi energy, $d\Omega_{\hat{\vec p}}$ is the differential solid angle at the position $\vec{p}$ on the Fermi surface, and $\widehat{g}^K(\vec{p},\varepsilon)$ is the Keldysh propagator, the trace of which determines the heat current response to a thermal gradient.
The formulation of the linear response of the chiral superfluid to an imposed temperature gradient, $\grad T$, including the interplay between branch conversion scattering and potential scattering by a random potential, is discussed in detail in Refs.~\cite{nga20}. 

\begin{figure}[t]
\begin{center}
\vspace*{-2cm}
\includegraphics[scale=0.5]{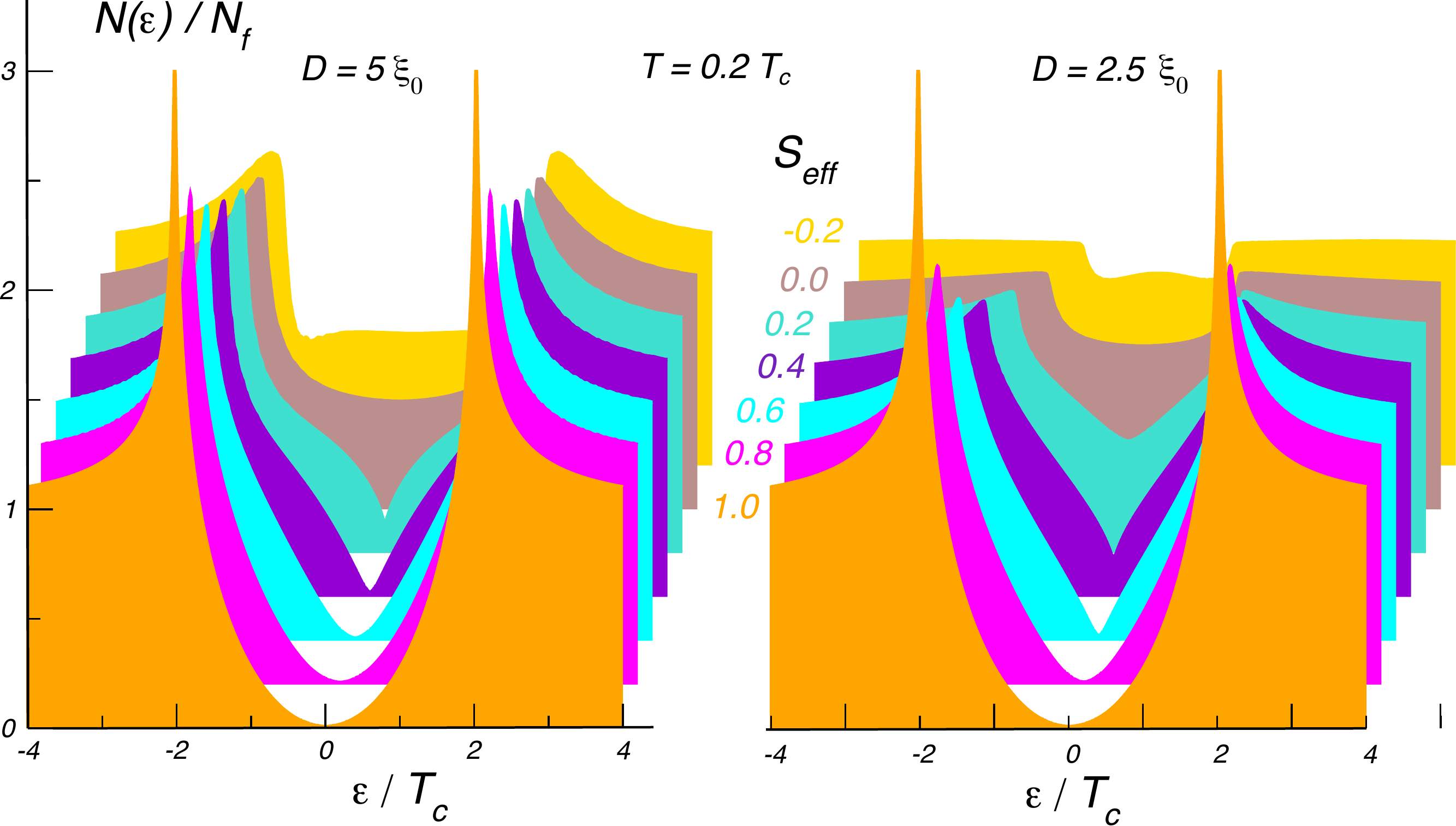}
\caption{DOS calculated for a slab of thickness $D = 5\,\xi_0 \sim 200\,{\rm nm}$ and $D = 2.5\,\xi_0 \sim 100\,{\rm nm}$ at $p=5.5\,{\rm bar}$ and $T = 0.2\,T_{c0}$ using surface boundary conditions in quasiclassical theory~\cite{vor03,hei21}.}
\label{fig-DOS_slabs}
\end{center}
\end{figure}

For superfluid $^3$He-A confined in aerogel, the latter is modeled as a homogeneous distribution of random impurities that scatter otherwise ballistically propagating $^3$He quasiparticles. This model is valid in the limit that the superfluid coherence length is sufficiently long compared to the typical distance between impurities. In this limit the disorder is parametrized by the mean-free path, $l$, which can be related to the aerogel density, or porosity, and transport cross-section of the impurities~\cite{thu98}. 
Within the transport theory for quasiparticles in the superfluid scattering of the random field is described in terms of the Nambu $\widehat{T}$-matrix that couples potential scattering by the random impurity potential to branch conversion scattering by the chiral order parameter that in general varies between the trajectory of the incident and scattered quasiparticle. The forward scattering limit of the $\widehat{T}$-matrix also determines the nonequilibrium corrections to the self-energy, $\widehat{\Sigma}^{K}_{{\rm imp}}(\vec{p},\varepsilon)=n_{{\rm imp}}\,\widehat{T}$, where $n_{{\rm imp}}$ is the mean density of impurities. 
The solutions for the propagator, $\widehat{g}^{K}$, the $\widehat{T}$-matrix and self-energy, $\widehat{\Sigma}^{K}_{{\rm imp}}$ have been derived in the linear response limit~\cite{gra96a,sha03,nga20,sha22}.
We use the linear response theory and calculations developed for thermal transport in bulk $^3$He-A confined in aerogel~\cite{sha22} to calculate the expected thermal Hall conductivity for $^3$He-A confined in a thin slab with $D\sim\xi$ and variable atomic scale roughness. 

\begin{figure}[t]
\begin{center}
\includegraphics[scale=0.7]{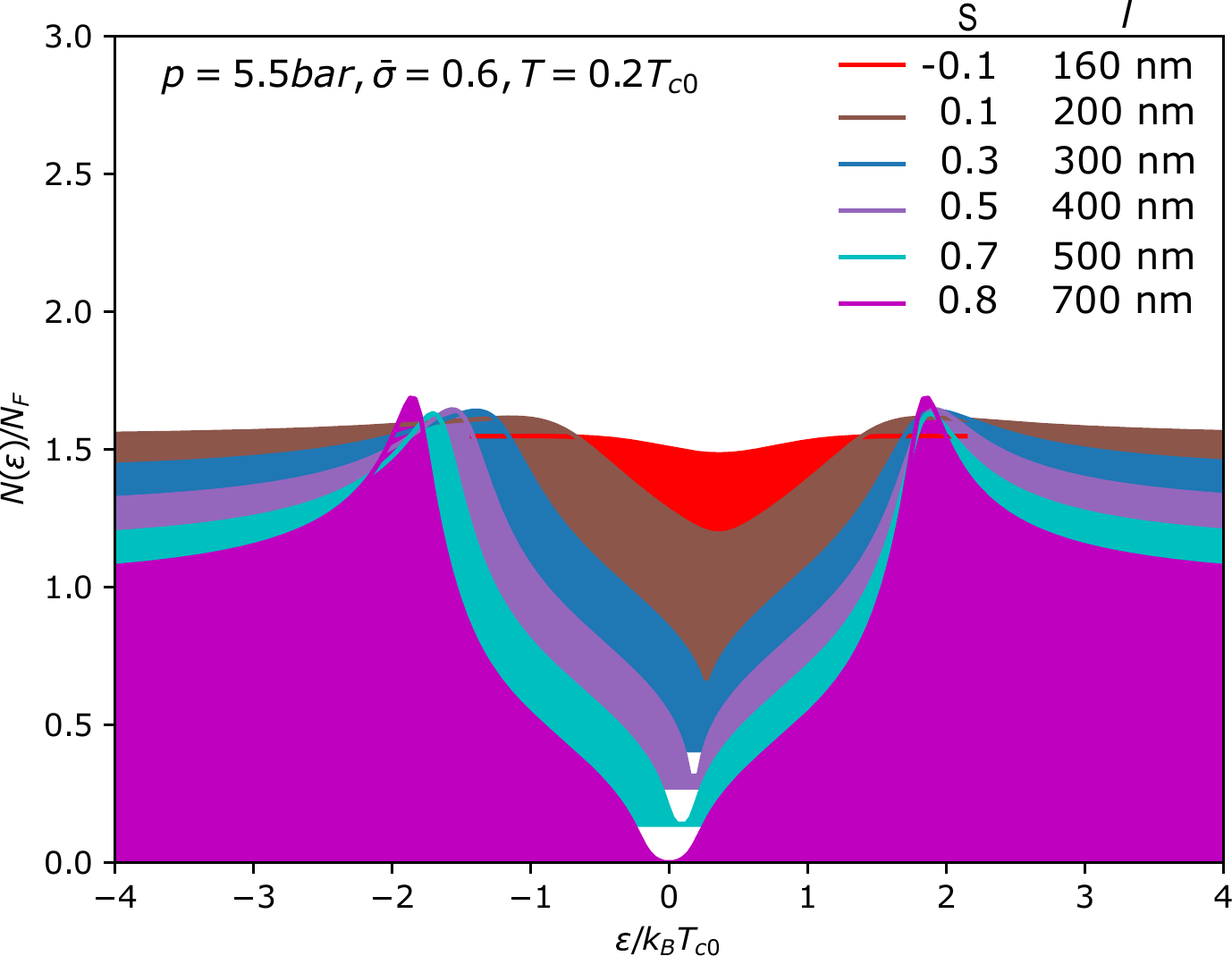}
\caption{Subgap DOS calculated for impurity disorder with $\bar\sigma = 0.6$ at $p=5.5\, {\rm bar}$ and $T = 0.2 T_{c0}$. The surface roughness parameter, $S$, for confined $^3$He-A in a slab of width $D=200\,{\rm nm}$ that yields the same sub-gap spectrum is paired with the mean-free path $l$.
}
\label{fig-DOS_aerogel-slab_mapping}
\end{center}
\end{figure}

\section{Aerogel and Slab Comparison}

The connection between $^3$He-A in aerogel and $^3$He-A confined in a slab with rough surfaces is made via the sub-gap quasiparticle spectrum calculated for the two systems. In particular, we compare the sub-gap spectrum for $^3$He-A in aerogel as a function of the mean-free path $l$ with the corresponding spectrum of $^3$He-A confined in a slab with rough boundaries parametrized by specularity parameter $-1\le S < 1$, where 
$S=1$ corresponds to specular scattering (no-pair-breaking for $^3$He-A),
$S=0$ corresponds to diffuse scattering (significant pair-breaking)~\cite{vor03}, 
and 
$S=-1$ for retro-reflecting surfaces (maximal pair-breaking)~\cite{sau11,hei21}.
For a slab of thickness $D=200\,{\rm nm}$ ($D=100\,{\rm nm}$) at $p=5.5\,{\rm bar}$ and $T=0.2\,T_{c_0}$, the calculated DOS is shown in the left (right) panel of Fig.~\ref{fig-DOS_slabs}. Note the increase in sub-gap states at low energies with increased surface roughness parametrized by $S$.

The comparison with the sub-gap spectrum of $^3$He-A in aerogel with mean-free-path $l$ provides a mapping of $l$ to $S$ based on equivalent sub-gap spectra. Our calculations for the longitudinal and transverse components of the thermal conductivity tensor for $^3$He-A in aerogel are then used to estimate the the conductivity tensor for $^3$He-A confined in a slab with surface roughness parameter $S$.
In Fig.~\ref{fig-DOS_aerogel-slab_mapping} we show the sub-gap DOS calculated for $^3$He-A embedded in a random impurity potential defined by mean free path $l$ and a normalized s-wave cross-section of $\bar\sigma=0.6$, as well as the corresponding surface roughness parameter $S$ for confined $^3$He-A confined in a $D=200\,{\rm nm}$ slab for the same pressure and temperature. 

In addition to broken mirror and time-reversal symmetry, particle-hole symmetry violation is required for a non-vanishing Hall conductivity~\cite{yip92}. The random impurity potential breaks the particle-hole symmetry of the pure superfluid ground state. The degree of particle-hole asymmetry depends on the scattering phase shift $\delta_l$ (for scattering in the $l$-wave channel), and thus the normalized cross-section $\bar\sigma=\sin^2\delta_0$, for s-wave impurities. Within the single channel s-wave scattering model particle-hole asymmetry vanishes in both the Born ($\bar\sigma\rightarrow 0$) and Unitarity ($\bar\sigma\rightarrow 1$) limits, and as a result the thermal Hall conductivity vanishes in these two limits. In the s-wave impurity model the Hall response is maximum for $\bar\sigma\sim 0.6$.

\begin{figure}[t]
\begin{center}
\includegraphics[scale=0.8]{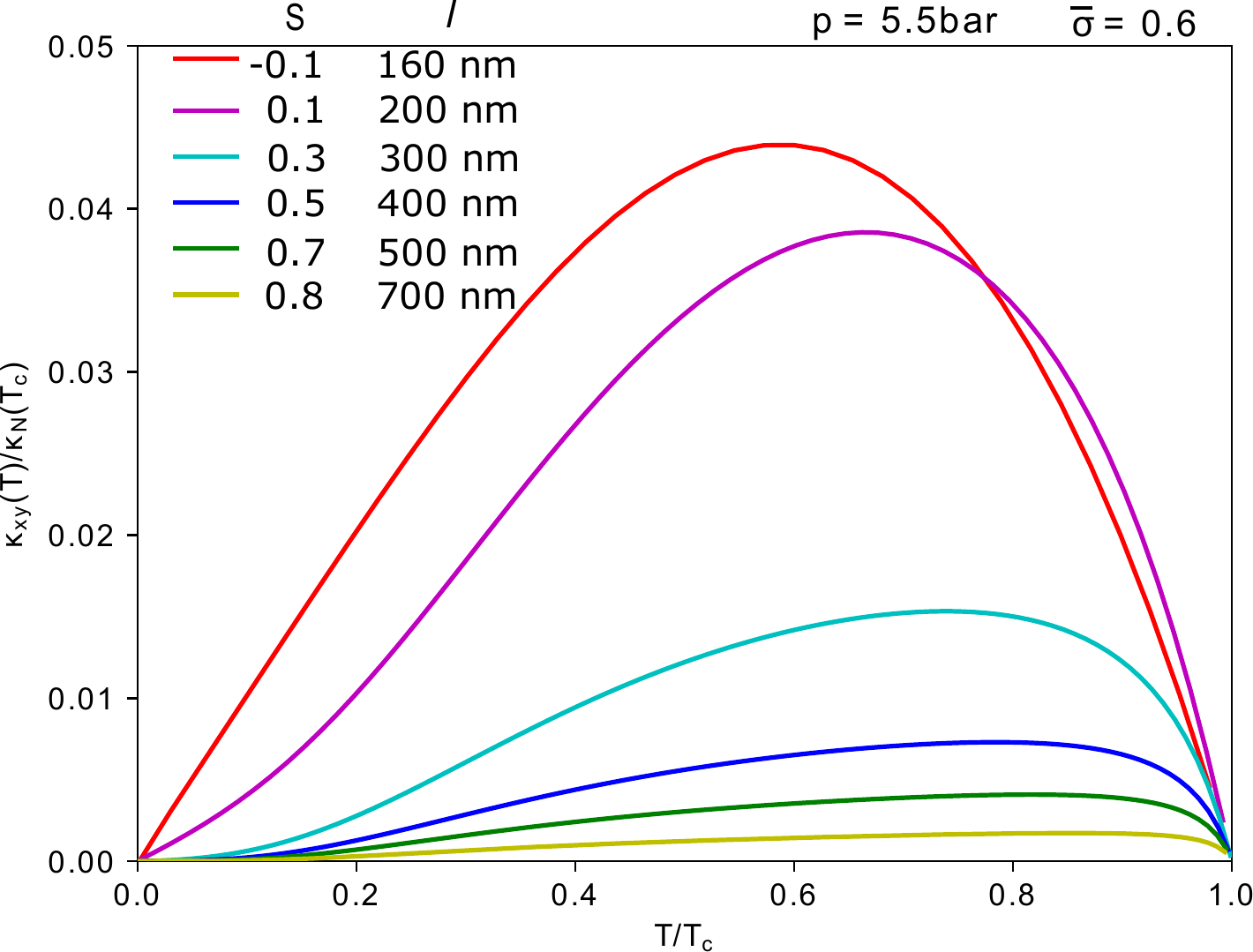}
\caption{Thermal Hall conductivity for various specularity parameters at $p=5.5 bar$. $\kappa_{xy} \rightarrow 0$ in the limit $S\rightarrow 1$ and for $S \le 0$. The maximum value of $\kappa_{xy}$ is obtained for $\bar\sigma=0.6$ shown here.}
\label{fig-Kxy_vs_S}
\end{center}
\end{figure}

The thermal Hall conductivity normalized to the normal state value of the longitudinal thermal conductivity at $T_c$ is shown in Fig.~\ref{fig-Kxy_vs_S} as a function of mean free path $l$ for $\bar\sigma=0.6$ with the corresponding surface roughness parameter, $S$, for $^3$He-A confined in a $D=200\,{\rm nm}$ slab.
Thus, the $\kappa_{xy}$ for confined $^3$He-A with rough surfaces is estimated to be a few percent of the normal-state thermal conductivity. We hope this result motivates experiments to measure the heat transport in slabs of the chiral phase of $^3$He which is known to be stabilised in sub-micron films. 
Observation of the thermal Hall transport would provide a definitive signature of the chirality of the confined superfluid phase. 
As the thermal Hall conductivity is sensitive to the degree of surface roughness, such experiments may also be used to better characterize the nature of surface scattering that dominates the physics of the superfluid under confinement.

\section*{Acknowledgments}

The research of JAS was supported by the National Science Foundation (Grant DMR-1508730). PS thanks Royal Holloway University of London for their affiliation. ABV acknowledges support by the National Science Foundation under Grant No. DMR-2023928.


\begin{thebibliography}{10}

\bibitem{pol08}
J~Pollanen, K~R Shirer, S~Blinstein, J~P Davis, H~Choi, T~M Lippman, W~P
  Halperin, and L~B Lurio.
{Globally anisotropic high porosity silica aerogels}.
{\em Journal Of Non-Crystalline Solids}, 354(40-41):4668--4674, 2008.

\bibitem{and61}
P.~W. Anderson and P.~Morel.
{Generalized Bardeen-Cooper-Schrieffer States and the Proposed
  Low-Temperature Phase of $^3$He}.
{\em Phys. Rev.}, 123:1911, 1961.

\bibitem{pol12}
J~Pollanen, J.~I.~A. Li, C.~A. Collett, W.~J. Gannon, W~P Halperin, and J~A
  Sauls.
{New chiral phases of superfluid $^3$He stabilized by anisotropic
  silica aerogel}.
{\em Nat. Phys.}, 8(4):317--320, 2012.

\bibitem{sha03}
P.~Sharma and J.~A. Sauls.
{Thermal Conductivity of Superfluid {$^3$He} in Aerogel}.
{\em Physica B}, 329-333:313--315, 2003.

\bibitem{kel65}
L.~V. Keldysh.
{Diagram Technique for Nonequilibrium Processes}.
{\em Zh. Eskp. Teor. Fiz.}, 47:1515, 1965.
English: Sov. Phys. JETP, 20, 1018 (1965).

\bibitem{eil68}
G.~Eilenberger.
{Transformation of Gorkov's Equation for Type II Superconductors into
  Transport-Like Equations}.
{\em Zeit. f. Physik}, 214:195, 1968.

\bibitem{lar69}
A.~I. Larkin and {Yu}.~N. Ovchinnikov.
{Quasiclassical Method in the Theory of Superconductivity}.
{\em Sov. Phys. JETP}, 28:1200--1205, 1969.

\bibitem{ser83}
J.~W. Serene and D.~Rainer.
{The Quasiclassical Approach to $^3He$}.
{\em Phys. Rep.}, 101:221, 1983.

\bibitem{nga20}
Vudtiwat Ngampruetikorn and J.~A Sauls.
{Impurity-induced Anomalous Thermal Hall Effect in Chiral
  Superconductors}.
{\em Phys. Rev. Lett.}, 124:157002, 2020.

\bibitem{vor03}
A.~Vorontsov and J.~A. Sauls.
{Thermodynamic Properties of Thin Films of Superfluid $^3$He-A}.
{\em Phys. Rev. B}, 68:064508, 2003.

\bibitem{hei21}
P.~J. Heikkinen, A.~Casey, L.~V. Levitin, X.~Rojas, A.~Vorontsov, P.~Sharma,
  N.~Zhelev, J.~M. Parpia, and J.~Saunders.
{Fragility of surface states in topological superfluid $^3$He}.
{\em Nat. Comm.}, 12:1574, 2021.

\bibitem{thu98}
E.~V. Thuneberg, S.-K. Yip, M.~Fogelstr\"om, and J.~A. Sauls.
{Scattering Models for Superfluid $^3$He in Aerogel}.
{\em Phys. Rev. Lett.}, 80:2861, 1998.

\bibitem{gra96a}
M.~J. Graf, S.-K. Yip, J.~A. Sauls, and D.~Rainer.
{Electronic Thermal Conductivity and the Wiedemann-Franz Law for
  Unconventional Superconductors}.
{\em Phys. Rev. B}, 53:15147, 1996.

\bibitem{sha22}
Priya Sharma and J.~A Sauls.
{Anomalous Thermal Hall Effect in Chiral Phases of {$^3$He}-Aerogel}.
{\em J. Low Temp. Phys.}, online: 1/8/2022:1--16, 2021.

\bibitem{sau11}
J.~A. Sauls.
{Surface states, Edge Currents, and the Angular Momentum of Chiral
  $p$-wave Superfluids}.
{\em Phys. Rev. B}, 84:214509, 2011.

\bibitem{yip92}
S.-K. Yip and J.~A Sauls.
{Circular Dichroism and Birefringence in Unconventional Superconductors},
{\em J. Low Temp. Phys.}, 86, 257-290, 1992.

\end{thebibliography}

\end{document}